**Experimental determination of the elastic cotunneling rate in a hybrid single-electron box**

Chia-Heng Sun[1], Po-Chen Tai[1], Jheng-An Jiang[2], Cen-Shawn Wu[2], Jeng-Chung Chen[3], Yung-Fu Chen[1]*

[1]Department of Physics, National Central University, Jhongli 32001, Taiwan

[2]Department of Physics, National Changhua University of Education, Changhua 500, Taiwan

[3]Department of Physics, National Tsing-Hua University, Hsinchu 30013, Taiwan

We report measurements of charge configurations and charge transfer dynamics in a hybrid single-electron box composed of aluminum and copper. We used two single-electron transistors (SETs) to simultaneously read out different parts of the box, enabling us to map out stability diagrams of the box and identify various charge transfer processes in the box. We further characterized the elastic cotunneling in the box, which is an important source of error in electron turnstiles consisting of hybrid SETs, and found that the rate was as low as 1 Hz at degeneracy and compatible with theoretical estimates for electron tunneling via virtual states in the central superconducting island of the box.

An ampere, i.e., the SI unit of current, is presently defined in terms of the force between two straight, parallel wires when an electric current is maintained;[1] however, this definition is indirect and not supported by precise quantum phenomena. Single-electron devices have been developed in which the enormous energy cost of adding an additional charge to a small conductor[2] enables individual electrons to be accurately manipulated. This development has naturally raised issues concerning the definition of the ampere. Over the past two decades, several attempts have been made to develop an ampere definition that is consistent with that of an electric current[3] by causing individual electrons to flow sequentially and synchronously through single-electron devices such as multi-island pumps[4, 5] and quantum-dot-based single-electron turnstiles.[6, 7] However, no realization to date has met all of the metrological requirements of the electric current standard in terms of accuracy and amplitude.[3]

Hybrid single-electron transistors (SETs), which consist of a normal metal (N) and a superconductor (S), have been demonstrated as a promising scheme for an electric current standard because of their simple design, easy operation, and remarkable accuracy.[8] An NISIN-SET electron turnstile in which one S island is linked to two N electrodes via two insulating barriers (I) has been demonstrated to reach 10 pA with a dc current, at an accuracy above $10^{-3}$.[8] However, several drawbacks in the NISIN system have been predicted,[9] particularly the inevitable error counts resulting from electron cotunneling; thus, the counterpart SINIS has been preferred for subsequent electron turnstile developments.[10, 11] Therefore, an experimental approach for determining the cotunneling rate is needed to investigate higher-order charge transfer processes in single-electron devices. In this letter, we describe an experimental method for determining the electron cotunneling rate in an NISIN system that is consistent with the prediction for electron tunneling via virtual states in an S island.[12]

We investigated a hybrid single-electron box named as an NISIN-box (see Fig. 1(b) for a simplified schematic), which consists of an S island connecting two N islands (N1 and N2) via two tunneling junctions to simulate an NISIN-SET electron turnstile. The box was electrically isolated and controlled by three adjacent gates. The charges could be transferred via tunneling among N1, S, and N2 but could not be transferred to the external environment. Variations in the excess charge numbers in N1 and N2 were monitored by two capacitively coupled SINIS-SETs.[13, 14] The $4 \times 10^{-4}\ e/\sqrt{Hz}$ charge sensitivity of our SETs was sufficient to identify charge transfers in the box with a single-electron resolution, and rates up to $10^3$ Hz could be determined. Note that the coupling capacitance $C_C$ between the SETs and the box reached the order of 10 aF, which is small compared to the gate capacitance $C_N$ and the tunnel junction capacitance $C_T$; thus, the SET measurements did not significantly perturb the box.

The NISIN-box and SINIS-SETs were fabricated by electron-beam lithography and shadow evaporation using 25 nm of aluminum (Al), followed by 35 nm of copper (Cu), on a thermally oxidized silicon substrate (see Fig. 1 (a) for a scanning electron micrograph (SEM) of the device). AlO$_x$ tunneling barriers between Al and Cu were formed by the thermal oxidation of Al in pure oxygen ($8 \times 10^{-2}$ torr) for 2 minutes immediately after Al deposition. The size of the S island in the NISIN-box was approximately $1000 \times 50 \times 25$ nm$^3$, and the N islands were much larger than the S island. The area and resistance of the tunneling junctions were approximately $150 \times 50$ nm$^2$ and 1 Mohm, respectively. Note that the quality and uniformity of the tunneling junctions are important because six junctions (two junctions for the box and four junctions for the SETs) are required to form the device. The measurements were performed using filtered leads in a dilution refrigerator with a base temperature of approximately 16 mK. The bias point of the box was set by three gate voltages, $V_{N1}$, $V_{N2}$, and $V_S$, and the charge state of the box was inspected using two SETs. Both SETs were voltage-biased near the edge of the blockade regime, where the SET currents exhibited large modulations, and the currents (on the order of 100 pA) were simultaneously measured using current preamplifiers and ADCs.

Figure 1(c) shows both SET currents ($I_{SET1}$ and $I_{SET2}$) as a function of $V_S$. The periodic oscillations represent the capacitive responses of the SETs to $V_S$. In addition to the periodic oscillations, several discontinuities were observed during the $V_S$ sweep due to charge transfer events occurring in the NISIN-box. For example, the discontinuity in $I_{SET1}$ marked by a blue arrow corresponds to a positive $V_S$ compensation for the SET1 island, which was required to maintain the same current level. This discontinuity implies that SET1 was affected by the decrease in potential at N1 due to the entrance of an electron. Similarly, the discontinuity in $I_{SET2}$ marked by a red arrow indicates the increase in potential at N2 caused by an electron exit. The charge transfer events occurring in the NISIN-box could therefore be recorded by continuously monitoring $I_{SET1}$ and $I_{SET2}$.

The charge configuration in the NISIN-box was manipulated using the gate voltages $V_{N1}$, $V_{N2}$, and $V_S$. Figure 2(a) shows $I_{SET1}$ (as a colormap) as a function of $V_{N2}$ and $V_S$ for $V_{N1} = 0$. The diagonal, colored, and equally spaced stripes show the periodic oscillations of the $I_{SET1}$ response to $V_{N2}$ and $V_S$. In addition, several discontinuities were observed and can be categorized as follows: i) as horizontal cuts with a period of 2.5 mV along $V_{N2}$ (highlighted by the rectangular box on the left and in Fig. 2(b)); ii) as lower left to upper right diagonal blurry bands across the plot (indicated by three arrows that are parallel to the blurry bands and highlighted by the middle box and in Fig. 2(c)), where alternate wide and narrow spacings occur between the blurry bands and the patterns repeat with 80-mV periods in $V_S$; and iii) as irregular switching events in the narrow regions between the blurry bands (highlighted by the box on the right and in Fig. 2(d)). Figures 2(f) and (g) also show $I_{SET1}$ and $I_{SET2}$ as a function of $V_{N2}$ and $V_{N1}$ for $V_S = 0$. Both plots exhibit saw-tooth-type stripes caused by regular discontinuities at the same places along the diagonal direction with a slope of 1 and a period of 2.5 mV in both the $V_{N1}$ and $V_{N2}$ directions.

The aforementioned discontinuity features are related to charge transfer events occurring in the box. Charge transfer occurs most frequently between two degenerate charge configurations. That is, these discontinuities correspond to boundaries among different charge configurations in the box. Consequently, measurements of both SETs were used to construct stability diagrams of the box. We constructed stability diagrams of the NISIN-box for comparison with the measured diagrams by evaluating the free energy $F$ for various charge configurations and found the lowest energy state at each bias point. In this electronic system, $F$ is the electrostatic energy stored in all of the capacitors, where the energy provided by the constant voltage sources is excluded, plus the quasiparticle excitation for an odd number of excess electrons in S ($n_S$):

$$F = \sum_i \frac{Q_i^2}{2C_i} - \sum_j Q_j V_j + E_S, \quad E_S = \begin{cases} \Delta & \text{for odd } n_S \\ 0 & \text{for even } n_S \end{cases}. \quad (1)$$

In Eq. (1), $Q_i$ is the charge stored in the capacitor $C_i$, $Q_j$ is the charge flowing through the voltage source $V_j$, and $\Delta$ is the superconducting energy gap. The values $C_{N1} = C_{N2} = 128$ aF, $C_S = 4$ aF, $C_{T2} = C_{T1} = 50$ aF, and $\Delta = 200$ μeV were used to calculate the stability diagrams shown in Fig. 2(e) and (h). The biases of the box are denoted by the gate charges ($n_{N1g} = C_{N1}V_{N1}/e$, $n_{N2g} = C_{N2}V_{N2}/e$, and $n_{Sg} = C_S V_S/e$) in units of $e$. The numbers of excess electrons in N1 and N2 are represented by $n_1$ and $n_2$, respectively. The total charge in the NISIN-box was conserved, and the box was presumably neutral, i.e., $n_1 + n_2 + n_S = 0$. Hence, $(n_1, n_2)$ were sufficient to label the lowest charge state in a region. Figure 2(e) shows a colored zone with the same $n_S = -(n_1 + n_2)$: the even $n_S$ regions are shown in red, and the odd $n_S$ regions are shown in blue. The calculated stability diagrams are periodic in the gate charges because the quadratic contributions to $F$ from the gate charges are associated with the corresponding excess charge numbers. Figure 2(e) shows that $n_1$ decreased by 1 and $n_2$ increased by 1 whenever $n_{N2g}$ increased by 2. Moreover, $n_S$ increased by 1 when $n_{Sg}$ increased by 1, and the patterns repeated whenever $n_{Sg}$ changed by 2. Figure 2(h) shows that $n_1$ decreased by 1 and $n_2$ increased by 1 whenever $n_{N1g}$ decreased by 2 or $n_{N2g}$ increased by 2. The edges of the charge configurations in Fig. 2(e) and (h) are similar to the discontinuities in Fig. 2(a) and (f-g), respectively. Figure 2(e) was juxtaposed with Fig. 2(a) to classify several charge transfer processes. First, the horizontal boundaries between $(i, j)$ and $(i-1, j+1)$ in Fig. 2(e) correspond to the horizontal cuts with a period of 2.5 mV along $V_{N2}$ in Fig. 2(a). As the bias moved across a horizontal cut, an electron shifted from N1 to N2 by elastic cotunneling. The diagonal cuts with a period of 2.5 mV along both $V_{N1}$ and $V_{N2}$ in Fig. 2(f) and (g) that are in line with the diagonal boundaries between $(i, j)$ and $(i-1, j+1)$ in Fig. 2(h) also indicate elastic cotunneling of electrons. Second, the three zigzag boundaries between successive $n_S$ values in Fig. 2(e) (highlighted by three arrows) correspond to the diagonal blurry bands in Fig. 2(a). As the bias moved across a blurry band along positive $V_S$, an electron was added to S through quasiparticle tunneling or Cooper-pair–electron cotunneling (a third-order process)[9], and $n_S$ increased by 1. The blurry appearance resulted from random switching and a broad transition between $n_S$ states differing by 1. The alternating spacings between the blurry bands resulted from the additional energy cost $\Delta$ for a single quasiparticle excitation for odd $n_S$ values; the wide spacing regions had fixed even $n_S$ states, and the narrow spacing regions had odd $n_S$ states. The alternating regions for even and odd $n_S$ states arose because the charging energy for an S island was larger than $\Delta$; thus, an Andreev reflection (a two-electron tunneling process) was energetically unfavorable and was not observed experimentally. Meanwhile, the charging energy for S was larger than those for N1 and N2; consequently, $n_S$ did not easily vary in contrast to $n_1$ and $n_2$, which is consistent with the circumstances in the NISIN-SET electron turnstile. Third, the random switchings in the odd $n_S$ regions (highlighted in Fig. 2(d)) originated from the inelastic cotunneling of electrons participated by an unpaired quasiparticle in S.[12]

We further characterized the elastic cotunneling events in the NISIN-box. Figure 3(a) shows a schematic of the elastic cotunneling process: an electron in N1 tunnels into S and forms a Cooper pair with another electron dissociating from a Cooper pair while the electron left behind tunnels into N2; the net result is that an electron moves from N1 to N2. Figure 3(b) shows $I_{SET1}$ and $I_{SET2}$ as a function of $V_{N1}$. Each sudden and simultaneous discontinuity marked by a pair of arrows indicates that N1 gained an electron and N2 lost an electron, i.e., electrons cotunneled from N2 to N1 as $V_{N1}$ increased. To observe the cotunneling dynamics, the device was biased near a degenerate point at which elastic cotunneling occurred (see Fig. 3(c)). The switching of $I_{SET1}$ between two positions indicates the occurrence of elastic cotunneling between N1 and N2 through S. Figure 3(d) shows six time traces of $I_{SET1}$ for a 100-Hz sampling rate at various values of $V_{N1}$. The random fluctuations of $I_{SET1}$ between the two states reveal the stochastic nature of the elastic cotunneling processes, whereas the two states with nearly equal populations indicate that the system was very close to degeneracy. As $V_{N1}$ increased, the cotunneling electron gradually stayed longer in N1, and the corresponding electron cotunneling rate out of N1, $\gamma_{el}$ (i.e., the inverse of the average lifetime for a cotunneling electron in N1), decreased. Figure 3(e) shows $\gamma_{el}$ out of N1 and N2 as a function of $V_{N1}$ as red triangles and blue circles, respectively.

The value of $\gamma_{el}$ in the NISIN-box was found to be as low as 1 Hz at degeneracy (corresponding to a NISIN-SET turnstile biased at zero voltage) and was on the order of 100 Hz away from the degeneracy (corresponding to a NISIN-SET turnstile biased near $\Delta/e$). To justify the measured value of $\gamma_{el}$, we calculated the electron cotunneling rate through a NISIN-SET using Eqs. (8) and (9) from Ref. 12. The lines in Fig. 3(e) correspond to the calculated $\gamma_{el}$. The calculated $\gamma_{el}$ qualitatively agrees with the measured values for the forward direction (electron cotunneling from a short-lived state to a long-lived state); in contrast, the measured $\gamma_{el}$ was much higher than the calculated value for the backward direction (cotunneling from a long-lived state to a short-lived state). The discrepancy may have resulted from photon-assisted tunneling caused by insufficient filtering or shielding of the measurement[15-17] and poor thermalization between the electrons and phonons in the box.[18] Nevertheless, the corresponding current error from cotunneling, $I_{el} = e\gamma_{el}$, was on the order of $10^{-17}$ amperes for a NISIN-SET turnstile biased near $\Delta/e$. When the turnstile was operated at 10 MHz, the corresponding accuracy was approximately $10^{-5}$, which still falls short of the $10^{-8}$ accuracy requirement for the ampere definition for metrology purposes by three orders of magnitude. It has been suggested that integrating a more dissipative environment into the system may reduce the errors from cotunneling;[19] we will explore this option in future studies.

In summary, the charge configurations and electron cotunneling dynamics in a NISIN-box were characterized by simultaneous measurements of two SETs. The qualitative behavior of the box near a degenerate point for cotunneling was consistent with the prediction for electron tunneling via virtual states in the S island, and the rate was as low as 1 Hz at degeneracy. Although the deductive errors from elastic cotunneling for NISIN-SET electron turnstiles operating at 10 MHz were three orders of magnitude above the desired precision, a hybrid-box monitored by two SETs was found, overall, to be a powerful tool for quantifying errors from higher-order processes in hybrid-SETs. In addition, other types of charge transfers in a NISIN-box, such as crossed Andreev reflections,[20-23] can be explored using the same measurement scheme.

The authors wish to thank V. Bubanja and C.-L. Lee for useful discussions. CHS, PCT, and YFC acknowledge support from the Ministry of Science and Technology in Taiwan under grant NSC-100-2112-M-008-017-MY3. This research was carried out in part at the Center for Nano Science and Technology, NCU.

Figure Captions:

Figure 1. (a) SEM of the NISIN-box readout using two SINIS-SETs. The darker metal is Al, and the lighter metal is Cu; the box is artificially colored red (S) and green (N), and the SETs are colored blue. The box is magnified in the inset, and the scale bar is 1 µm. (b) Schematic of the measurement, showing that the bias point of the box is determined by applying three voltages to the gates ($V_{N1}$, $V_{N2}$, $V_S$) and the charge state of the box is inspected using two capacitively coupled SETs. (c) SET currents ($I_{SET1}$ and $I_{SET2}$) as a function of $V_S$; $I_{SET2}$ is offset by -170 pA for clarity. The discontinuities in $I_{SET1}$ and $I_{SET2}$ result from modifications of the potential at N1 and N2 due to charge transfer events in the box.

Figure 2. Stability diagrams for the NISIN-box. (a) Measured $I_{SET1}$ as a function of $V_S$ and $V_{N2}$ for $V_{N1} = 0$. Horizontal and periodic discontinuities from elastic cotunneling are highlighted in the rectangular box on the left. Three diagonal blurry bands across the plot caused by quasi-particle tunneling are marked by three arrows parallel to the bands and are highlighted in the middle box; they are also evident in the inset, where high-pass filtered $I_{SET}$ is shown. Emphasize Random switching events originating from inelastic cotunneling of electrons due to an unpaired quasiparticle in S are highlighted in the box on the right. (b-d) Magnified images of the three individual boxes from left to right in (a), respectively, where the axis labels and units are the same as those in (a). (e) Calculated stability diagrams using $C_{N1} = C_{N2} = 128$ aF, $C_S = 4$ aF, and $C_{T1} = C_{T2} = 50$ aF to mimic the measured stability diagrams shown in (a); the axes indicate the gate charges in units

of $e$. The label ($n_1$, $n_2$) denotes the lowest charge state in the region. The red and blue areas indicate even and odd $n_S$ respectively. The three zigzag boundaries between successive $n_S$ states are highlighted by three parallel arrows corresponding to the three diagonal blurry bands marked by arrows in (a). Inset: high-pass filtered $I_{SET1}$ for wider $V_S$ and $V_{N2}$ spans. The patterns of the blurry bands repeat in $V_S$ and the spacings between the blurry bands depend on the parity of $n_S$. (f-g) Measured $I_{SET1}$ and $I_{SET2}$ as a function of $V_{N1}$ and $V_{N2}$ for $V_S = 0$; periodic discontinuities along the diagonal direction with a slope of 1 and a period of 2.5 mV in $V_{N1}$ and $V_{N2}$ axes are clearly shown in both plots at the same place and result from elastic cotunneling in the box, as detected by both SETs. (h) Calculated stability diagram to imitate the measured stability diagrams shown in (f) and (g).

Figure 3. Elastic cotunneling in the NISIN-box. (a) A schematic of elastic cotunneling from N1 to N2 through S. (b) $I_{SET1}$ and $I_{SET2}$ as a function of $V_{N1}$, where $I_{SET2}$ is negatively offset for clarity. The periodic oscillations of $I_{SET1}$ and $I_{SET2}$ show direct capacitive responses to Gate N1, and the smaller period in $I_{SET1}$ indicates that SET1 has stronger capacitive coupling to Gate N1 than SET2. The consecutive current discontinuities marked by arrows result from elastic cotunneling from N2 to N1 for increasing $V_{N1}$. (c) Detailed measurement of $I_{SET1}$ as a function of $V_{N1}$ near a degenerate point. The switching of $I_{SET1}$ between two positions indicates a cotunneling electron moving between N1 and N2. (d) Time traces of $I_{SET1}$ for various $V_{N1}$ values near a degenerate point. (e) Elastic cotunneling rate $\gamma_{el}$ as a function of $V_{N1}$. The symbols and lines show measured and calculated values of $\gamma_{el}$, respectively; red and blue denote

tunneling out of N1 and N2, respectively. The measured $\gamma_{el}$ is determined from the reciprocal of the average lifetime in a state. The calculated $\gamma_{el}$ value is based on Eqs. (8) and (9) from Ref. 12, with $E_1 = E_2 = 800$ µeV, $T = 50$ mK, and $G_n = 1.1 \times 10^{-6}$ µS.

Figure 1

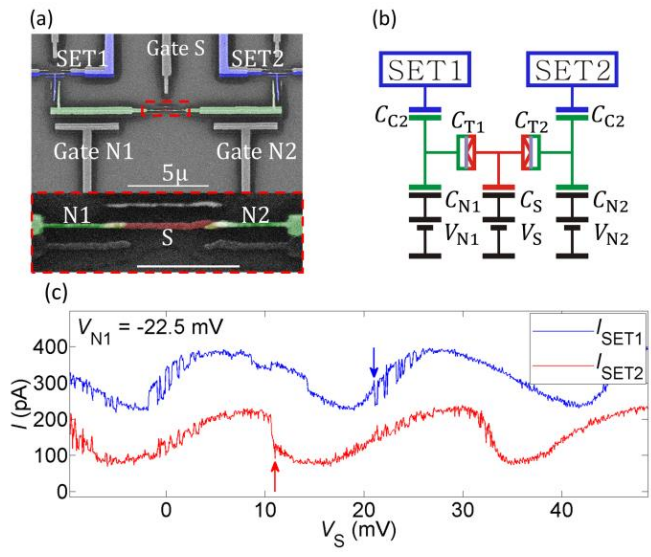

Figure 2

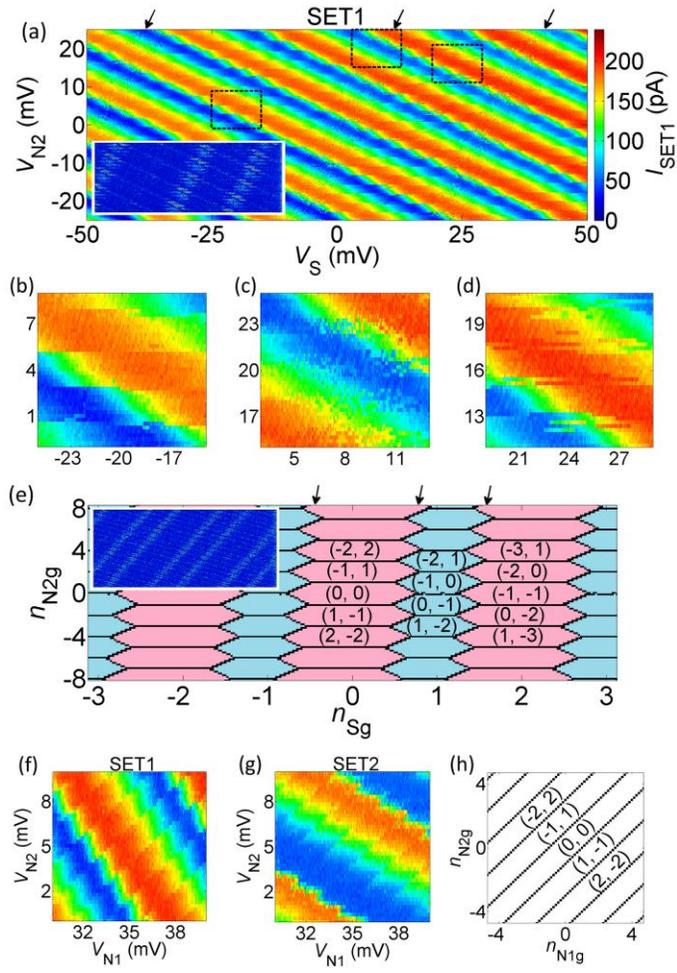

Figure 3

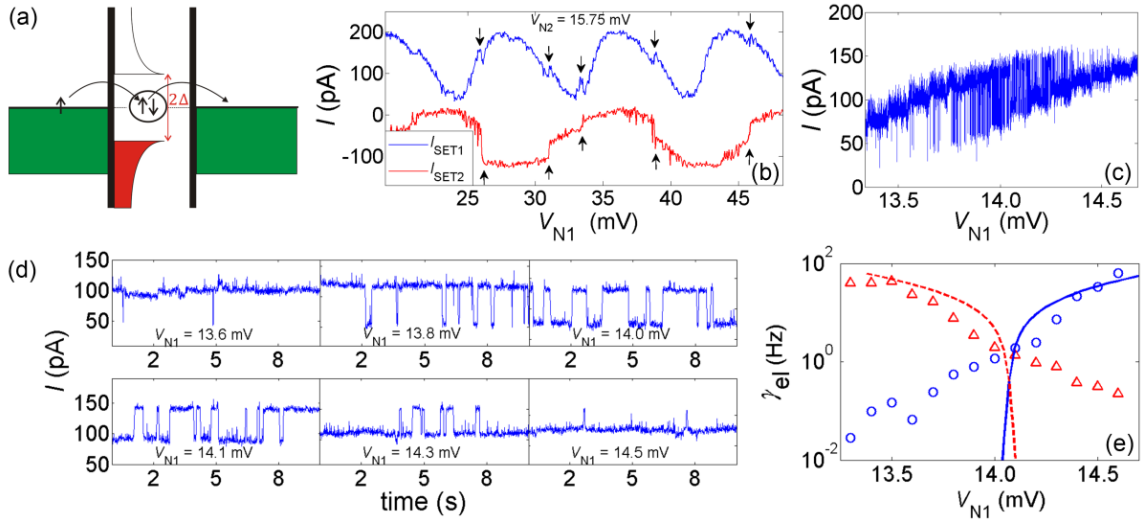